\documentclass[aps,showpacs,superscriptaddress,preprint]{revtex4}%
\usepackage{amsfonts}
\usepackage{amsmath}
\usepackage{amssymb}
\usepackage{graphicx}%
\providecommand{\U}[1]{\protect\rule{.1in}{.1in}}

\begin{document}
\title{First-principles calculations of phase transition, low elastic modulus, and superconductivity for zirconium}
\author{Bao-Tian Wang}
\thanks{E-mail: wbt11129@sxu.edu.cn}
\affiliation{Institute of Theoretical Physics and Department of
Physics, Shanxi University, Taiyuan 030006, People's Republic of
China} \affiliation{LCP, Institute of Applied Physics and
Computational Mathematics, P.O. Box 8009, Beijing 100088, People's
Republic of China}
\author{Peng Zhang}
\affiliation{Department of Nuclear Science and Technology, Xi'an
Jiaotong University, Xi'an 710049, People's Republic of China}
\author{Han-Yu Liu}
\affiliation{State Key Lab of Superhard Materials, Jilin University,
Changchun 130012, People's Republic of China}
\author{Wei-Dong Li}
\affiliation{Institute of Theoretical Physics and Department of
Physics, Shanxi University, Taiyuan 030006, People's Republic of
China}
\author{Ping Zhang}
\thanks{Corresponding author; zhang\_ping@iapcm.ac.cn}
\affiliation{LCP, Institute of Applied Physics and Computational
Mathematics, P.O. Box 8009, Beijing 100088, People's Republic of
China} \affiliation{Center for Applied Physics and Technology,
Peking University, Beijing 100871, People's Republic of China}

\pacs{62.20.Dc, 64.70.kd, 61.50.Ks, 74.70.Ad}

\begin{abstract}
The elasticity, dynamic properties, and superconductivity of
$\alpha$, $\omega$, and $\beta$ Zr are investigated by using
first-principles methods. Our calculated elastic constants, elastic
moduli, and Debye temperatures of $\alpha$ and $\omega$ phases are
in excellent agreement with experiments. Electron-phonon coupling
constant $\lambda$ and electronic density of states at the Fermi
level $N$(\emph{E}$_{\rm{F}}$) are found to increase with pressure
for these two hexagonal structures. For cubic $\beta$ phase, the
critical pressure for mechanical stability is predicted to be 3.13
GPa and at \emph{P}=4 GPa the low elastic modulus ($E$=31.97 GPa)
can be obtained. Besides, the critical pressure for dynamic
stability of $\beta$ phase is achieved by phonon dispersion
calculations to be $\mathtt{\sim}$26 GPa. Over this pressure,
$\lambda$ and $N$(\emph{E}$_{\rm{F}}$) of $\beta$ phase decrease
upon further compression. Our calculations show that the large value
of superconducting transition temperature $\emph{T}_{\rm{c}}$ at 30
GPa for $\beta$ Zr is mainly due to the TA1 soft mode. Under further
compression, the soft vibrational mode will gradually fade away.

\end{abstract}
\maketitle
\section{INTRODUCTION}
Group IV transition metals such as titanium, zirconium, and hafnium
have wide applications in the aerospace, nuclear, and biomedical
industry \cite{Ikehata,LiuPRB,Perez}. The body-centered cubic (bcc)
type ($\beta$ phase) of Ti and Zr alloys in usage as metallic
biomaterial have been studied extensively because of their
properties of nontoxicity, high strength, good biocompatibility, and
low elastic modulus \cite{Ikehata,HuAPL}. However, the
investigations of the single crystal $\beta$ Zr as biomaterial has
not been reported.

Actually, at ambient condition Zr crystallizes in hexagonal
closed-packed (hcp) structure ($\alpha$ phase). And at high
temperature of 1135 K, it transforms into the $\beta$ phase. At room
temperature and under pressure range of 2-7 GPa
\cite{Sikka,Xia1,ZhaoPRB}, the $\alpha$ phase transforms to another
more open hexagonal structure of $\omega$ phase with space group
\emph{P\rm{6}/mmm} (No. 191). And under further high pressure of
30-35 GPa, the $\omega$$\mathtt{\rightarrow}$$\beta$ phase
transition has been observed
\cite{Xia1,Xia2,Akahama1,Akahama2,ZhaoAPL}. The Young's moduli of
both $\alpha$ and $\omega$ phases at ambient condition are largely
bigger than that of a human bone (about 30 GPa). So the condition of
low elastic modulus cannot be satisfied for these two phases. As for
$\beta$ phase, it cannot be exploited because the reverse phase
transformations take place upon unloading. So it seems that we
should not consider metal Zr as candidate biomaterial. But,
fortunately recent experiment has successfully stabilized the
$\beta$ Zr at room temperature using compression stresses higher
than 3 GPa plus the controlled application of shear \cite{Perez}.
This makes stabilizing the high temperature or pressure phases of Zr
at ambient conditions reality. Therefore, the study of the elastic
modulus for $\beta$ Zr at low pressure is valuable.

As another key objective of our work, we calculate the
superconducting transition temperature $T$$_{c}$ for metal Zr.
Although the $T$$_{c}$ of Zr has been observed in experiments
\cite{Akahama1}, to date no theoretical work is found in the
literature. Our calculations will show that the increase (decrease)
in $T$$_{c}$ with pressure can be understood in terms of the
enhancement (reduction) of the electronic density of states (DOS) at
the Fermi level and the corresponding behaviors of electron-phonon
coupling constant. In present study, the dynamic stability of
$\beta$ Zr is tested carefully by performing phonon dispersion
calculations. The $\beta$$\mathtt{\rightarrow}$$\omega$ and
$\beta$$\mathtt{\rightarrow}$$\alpha$ transitions are discussed in
detail. The rest of this paper is arranged as follows. In Sec. II
the computational methods are described. In Sec. III we present and
discuss our results. In Sec. IV we summarize the conclusions of this
work.

\section{computational methods}
\subsection{Computational details}
The first-principles density functional theory (DFT) calculations on
the basis of the frozen-core projected augmented wave (PAW) method
of Bl\"{o}chl \cite{PAW} are performed within the Vienna \textit{ab
initio} simulation package (VASP) \cite{Kresse3}, where the Perdew,
Burke, and Ernzerhof (PBE) \cite{PBE} form of the generalized
gradient approximation (GGA) is employed to describe electron
exchange and correlation. For the plane-wave set, a cutoff energy of
500 eV is used. The $\Gamma$-centered \emph{k} point-meshes in the
full wedge of the Brillouin zone (BZ) are sampled by 18$\times
$18$\times$16, 16$\times$16$\times$18, and 18$\times$18$\times$18
grids according to the Monkhorst-Pack (MP) \cite{Monk} for $\alpha$
(two-atom cell), $\omega$ (three-atom cell), and $\beta$ (two-atom
cell) Zr, respectively, and all atoms are fully relaxed until the
Hellmann-Feynman forces become less than 0.001 eV/\AA. The Zr
4$d$$^{3}$5$s$$^{1}$ orbitals are explicitly included as valence
electrons. The pseudopotential plane-wave method within the PBE-GGA
through the QUANTUM-ESPRESSO package \cite{Baroni2} is employed to
calculate the electronic properties, lattice dynamics, and
electron-phonon coupling (EPC) for Zr, where the Zr
4$s$$^{2}$4$p$$^{6}$4$d$$^{2}$5$s$$^{2}$ are treated as valence
electrons. Convergence tests give the choice of kinetic energy
cutoffs of 60 Ry with Gaussians width of 0.05 Ry for all three
phases of Zr and 12$\times$12$\times$8, 8$\times$8$\times$12, and
16$\times $16$\times$16 MP grids of \emph{k}-point meshes for
$\alpha$, $\omega$, and $\beta$ Zr, respectively. Phonon frequencies
are calculated based on the density functional linear-response
method \cite{Baroni1,Giannozzi}. In the interpolation of the force
constants for the phonon dispersion curve calculations,
6$\times$6$\times$4, 4$\times$4$\times$6, and 8$\times $8$\times$8
\emph{q}-point meshes and denser 24$\times$24$\times$16,
16$\times$16$\times$24, and 32$\times $32$\times$32 \emph{k}-point
meshes in the first BZ are used for $\alpha$, $\omega$, and $\beta$
phases of Zr, respectively.

\subsection{Mechanical properties}
To avoid the Pulay stress problem, the geometry optimization at each
volume is performed at fixed volume rather than constant pressure.
The theoretical equilibrium volume, bulk modulus \emph{B}, and
pressure derivative of the bulk modulus \emph{B$^{\prime}$} are
obtained by fitting the energy-volume data in the third-order
Birch-Murnaghan equation of state (EOS) \cite{Birch}. Elastic
constants for cubic symmetry ($C_{11}$, $C_{12}$, and $C_{44}$) and
hexagonal structure ($C_{11}$, $C_{12}$, $C_{13}$, $C_{33}$, and
$C_{44}$) are calculated by applying stress tensors with various
small strains onto the equilibrium structures. The strain amplitude
$\delta$ is varied in steps of 0.006 from $\delta$=$-$0.036 to
0.036. After obtaining elastic constants, the polycrystalline bulk
modulus \emph{B} and shear modulus \emph{G} are calculated from the
Voigt-Reuss-Hill (VRH) approximations \cite{Hill}. The Young's
modulus \emph{E} and Poisson's ratio $\upsilon$ are calculated
through $E=9BG/(3B+G)$ and $\upsilon=(3B-2G)/[2(3B+G)]$. In
calculation of the Debye temperature ($\theta_{D}$), we use the
relation
\begin{align}
\theta_{D}=\frac{h}{k_{B}}\left( \frac{3n}{4\pi\Omega}\right)
^{1/3}\upsilon_{m},
\end{align}
where \emph{h} and $\emph{k}_{B}$
are Planck and Boltzmann constants, respectively, \emph{n} is the
number of atoms in the molecule, $\Omega$ is molecular volume, and
$\upsilon_{m}$ is the average sound wave velocity. The average wave
velocity in the polycrystalline materials is approximately given as
\begin{align}
\upsilon_{m}=\left[ \frac{1}{3}\left(
\frac{2}{\upsilon_{t}^{3}}+\frac {1}{\upsilon_{l}^{3}}\right)
\right]  ^{-1/3},\end{align} where $\upsilon_{t}=\sqrt{G/\rho}$
($\rho$ is the density) and $\upsilon_{l}=\sqrt{(3B+4G)/3\rho}$ are
the transverse and longitudinal elastic wave velocity of the
polycrystalline materials, respectively.

\subsection{Superconduntivity}
The superconducting transition temperature $T$$_{c}$ is evaluated by
using the Allen-Dynes modified McMillan equation
\cite{Allen,McMillan}
\begin{align}
T_{c}=\frac{\omega\rm{_{log}}}{1.2}\rm{exp}\left[
-\frac{1.04(1+\lambda)}{\lambda-\mu^{*}(1+0.62\lambda)}\right],
\end{align}
where
\begin{align}
\omega\rm{_{log}}=\rm{exp}\left[
\frac{2}{\lambda}\int_{0}^{\infty}\frac{\emph{d}\omega}{\omega}\alpha^{2}\emph{F}(\omega)\rm{ln}\omega\right]
\end{align}
is the logarithmic average frequency,
\begin{align}
\lambda=2\int_{0}^{\infty}\frac{\alpha^{2}\emph{F}(\omega)}{\omega}d\omega
\end{align}
is the electron-phonon coupling constant, and $\mu^{*}$ is the
Coulomb pseudopotential. The Eliashberg electron-phonon spectral
function $\alpha^{2}F(\omega)$ is written as
\begin{align}
\alpha^{2}F(\omega)=\frac{1}{2\pi\!N(E_{\rm{F}})}\sum_{qv}\frac{\gamma_{qv}}{\omega_{qv}}\delta(\omega-\omega_{qv}),
\end{align}
where $N$(\emph{E}$_{\rm{F}}$) is the electronic density of state
(DOS) at the Fermi level.

\section{results}
\begin{figure}[ptb]
\begin{center}
\includegraphics[width=0.5\linewidth]{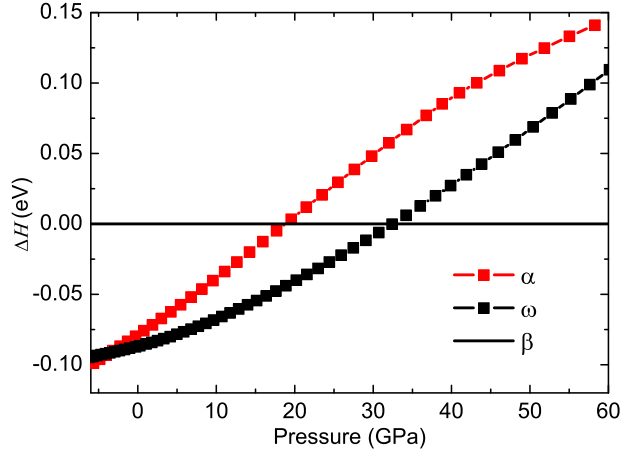}
\end{center}
\caption{(Color online) Calculated enthalpy differences of $\alpha$
and $\omega$
phases with respect to $\beta$ phase as functions of pressure.}%
\label{enthalpy}%
\end{figure}

\begin{table*}[ptb]
\caption{Calculated lattice parameters (\emph{a} or \emph{c}), bulk
modulus (\emph{B}), pressure derivative of the bulk modulus
(\emph{B$^{^{\prime}}$}), and elastic constants of $\alpha$,
$\omega$, and $\beta$ Zr at different pressures. For comparison,
experimental values and other theoretical results are also listed.}%
\label{elastic}
\begin{ruledtabular}
\begin{tabular}{cccccccccccccccccc}
Phase&Method&Pressure&a&c&\emph{B}&\emph{B$^{^{\prime}}$}&\emph{C$_{11}$}&\emph{C$_{12}$}&\emph{C$_{13}$}&\emph{C$_{33}$}&\emph{C$_{44}$}\\
&&(GPa)&({\AA})&({\AA})&(GPa)&&(GPa)&(GPa)&(GPa)&(GPa)&(GPa)\\
\hline
$\alpha$&This work&0&3.236&5.168&96.0&3.12&146.7&68.5&71.0&163.3&26.0\\
&&5.35&3.178&5.093&&&162.3&78.4&81.9&187.5&24.9\\
&DFT-PBE$^{\emph{a}}$&0&3.240&5.178&93.4&3.22&141.1&67.6&64.3&166.9&25.8\\
&DFT-PBE$^{\emph{b}}$&0&3.232&5.182&93.4&3.22&139.4&71.3&66.3&162.7&25.5\\
&Expt.&0&3.233$^{\emph{c}}$&5.146$^{\emph{c}}$&92$^{\emph{c}}$&4.0$^{\emph{c}}$&144.0$^{\emph{d}}$&74.0$^{\emph{d}}$&67.0$^{\emph{d}}$&166.0$^{\emph{d}}$&33.0$^{\emph{d}}$\\
$\omega$&This work&0&5.036&3.152&96.9&3.39&161.7&72.6&53.5&195.6&33.7\\
&&6.04&4.939&3.097&&&187.0&85.8&63.1&224.4&37.3\\
&&10.01&4.884&3.066&&&201.5&94.5&69.0&241.1&38.8\\
&DFT-PBE$^{\emph{a}}$&0&5.056&3.150&101.1&3.27&165.2&75.6&47.5&198.7&30.6\\
&Expt.&0&5.039$^{\emph{e}}$&3.150$^{\emph{e}}$&104.0$^{\emph{f}}$&2.8$^{\emph{f}}$&&&&&\\
$\beta$&This work&0&3.574&&90.2&3.06&86.6&92.3&&&26.6\\
&&10&3.461&&&&123.1&106.6&&&32.0\\
&&20&3.374&&&&160.0&116.8&&&37.0\\
&&30&3.302&&&&196.9&124.1&&&42.2\\
&&35&3.271&&&&216.7&124.4&&&45.8\\
&&40&3.241&&&&235.2&128.4&&&49.2\\
&&50&3.188&&&&265.6&136.2&&&54.4\\
&&60&3.141&&&&298.3&141.3&&&61.7\\
&DFT-PBE$^{\emph{b}}$&0&3.580&&&&84.2&91.4&&&32.3\\
&LDA$^{\emph{g}}$&0&&&&&&&&&32.8\\
&Expt.&0&3.627$^{\emph{c}}$&&66$^{\emph{c}}$&&104$^{\emph{h}}$&93$^{\emph{h}}$&&&38$^{\emph{h}}$\\
\end{tabular}
$^{\emph{a}}$ Reference \cite{HaoJPCM}, $^{\emph{b}}$ Reference
\cite{Ikehata}, $^{\emph{c}}$ Reference \cite{ZhaoPRB},
$^{\emph{d}}$ Reference \cite{Brandes},  $^{\emph{e}}$ Reference
\cite{Barrett}, $^{\emph{f}}$ Reference \cite{LiuPRB}, $^{\emph{g}}$
Reference \cite{Ahuja}, $^{\emph{h}}$ Results measured at
$T$=915$^{\circ}$C from reference \cite{Heiming}.
\end{ruledtabular}
\end{table*}
At 0 K, the Gibbs free energy is equal to the enthalpy \emph{H}.
After calculation, we can plot in Fig. \ref{enthalpy} the relative
enthalpies of the $\alpha$ and the $\omega$ phases with respect to
the $\beta$ phase as functions of pressure. The crossing between the
$\alpha$ and $\omega$ enthalpy curves readily gives phase transition
pressure of $-$3.7 GPa, which indicates that at ambient pressure the
$\omega$ phase is more stable than the $\alpha$ phase. This fact is
in disagreement with experiment, but coincides well with recent
DFT-PBE \cite{HaoPRB} and FP-LAPW \cite{Schnell} calculations. The
disagreement between theory and experiment mainly originates from
temperature effect. Actually, previous DFT-PBE \cite{HaoPRB} study
clearly shows that the transition of
$\alpha$$\mathtt{\rightarrow}$$\omega$ occurs at 1.7 GPa at
\emph{T}=300 K. As for the $\omega$$\mathtt{\rightarrow}$$\beta$
transition, a pressure of 32.4 GPa is obtained. This value falls
exactly in the pressure range of 30$-$35 GPa measured by experiments
\cite{Xia1,Xia2,Akahama1,Akahama2,ZhaoAPL} and also fairly accords
with other calculations \cite{HaoPRB,Schnell}.

\begin{figure}[ptb]
\begin{center}
\includegraphics[width=0.5\linewidth]{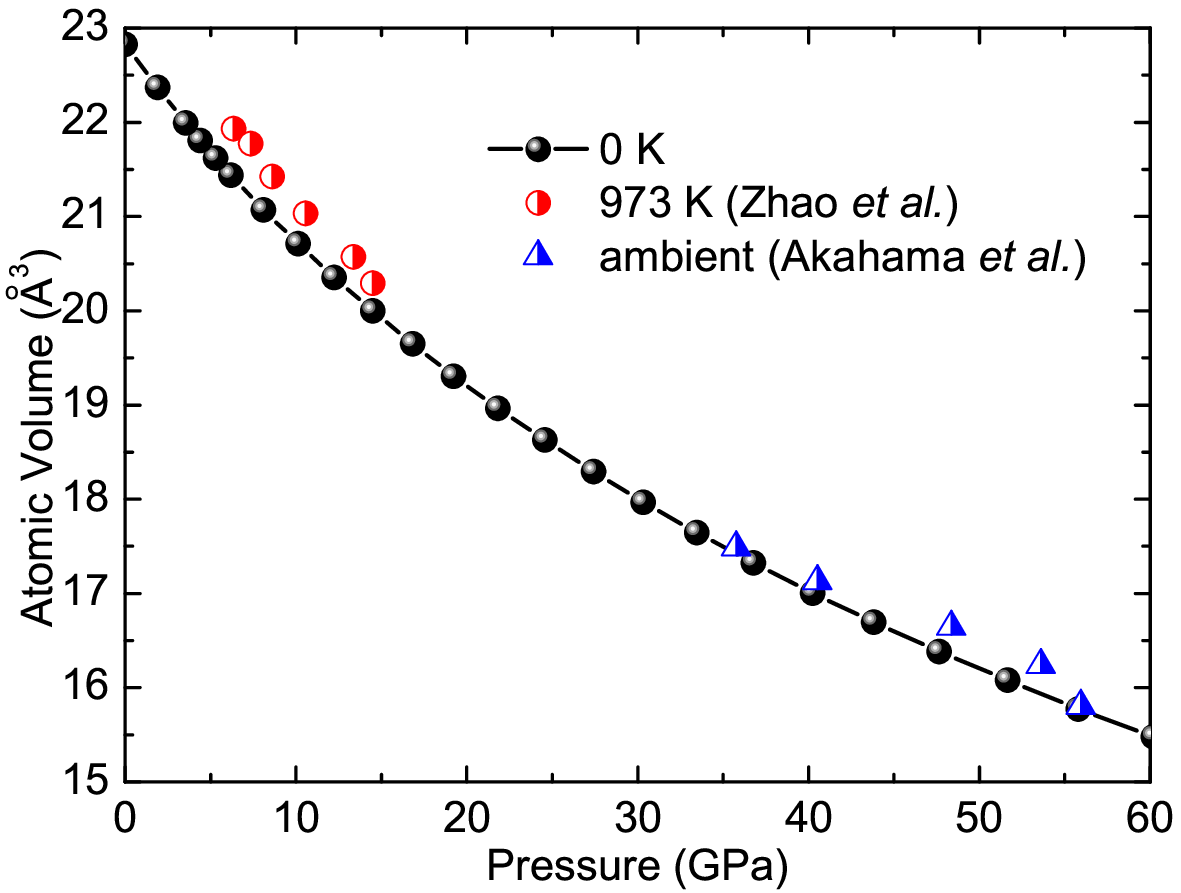}
\end{center}
\caption{(Color online) Calculated compression curves of $\beta$ Zr
compared with experimental measurements.}%
\label{compression}%
\end{figure}

\begin{table*}[ptb]
\caption{Calculated elastic moduli, Poisson's ratio ($\upsilon$),
density ($\rho$), transverse ($\upsilon_{t}$), longitudinal
($\upsilon_{l}$), and average ($\upsilon_{m}$) sound velocities
calculated from polycrystalline bulk and shear modulus, and Debye
temperature of $\alpha$, $\omega$, and $\beta$ Zr at different
pressures. For comparison,
experimental values and other theoretical results are also listed.}%
\label{mechanical}
\begin{ruledtabular}
\begin{tabular}{cccccccccccccccccc}
Phase&Method&Pressure&\emph{B}&\emph{G}&\emph{E}&$\upsilon$&$\rho$&$\upsilon_{t}$&$\upsilon_{l}$&$\upsilon_{m}$&$\theta_{D}$\\
&&(GPa)&(GPa)&(GPa)&(GPa)&&(g/cm$^{3}$)&(km/s)&(km/s)&(km/s)&(K)\\
\hline
$\alpha$&This work&0&97.4&33.8&91.0&0.344&6.465&2.288&4.695&2.571&267.3\\
&&5.35&110.4&34.9&94.8&0.357&6.804&2.266&4.803&2.550&269.8\\
&Expt.$^{\emph{a}}$&0&95.3&36.3&96.6&0.331&6.511&2.361&4.698&&275.9&&\\
$\omega$&This work&0&97.5&43.6&113.7&0.306&6.568&2.575&4.868&2.878&300.9\\
&&6.04&113.5&49.0&128.5&0.311&6.948&2.655&5.074&2.970&276.4\\
&&10.01&123.2&51.6&135.8&0.316&7.177&2.681&5.172&3.001&282.3\\
&Expt.$^{\emph{a}}$&0&104.0&45.1&118.3&0.311&6.589&2.616&4.996&&306.1\\
&&6.88&&&&&&2.638&5.177&&\\
&&10.59&&&&&&2.652&5.284&&\\
$\beta$&This work&10&112.1&18.7&53.1&0.421&7.309&1.599&4.330&1.816&196.7\\
&&20&131.2&29.8&83.2&0.394&7.889&1.944&4.655&2.199&244.4\\
&&30&148.4&39.8&109.6&0.377&8.415&2.175&4.893&2.454&278.7\\
&&35&156.5&45.5&124.6&0.367&8.662&2.293&5.008&2.584&296.3\\
&&40&164.0&50.9&138.3&0.360&8.900&2.390&5.104&2.691&311.4\\
&&50&179.3&58.3&157.8&0.353&9.353&2.497&5.243&2.809&330.4\\
&&60&193.6&68.0&182.5&0.343&9.781&2.636&5.391&2.961&353.5\\
&Expt.$^{\emph{b}}$&0&96.7&18.1&51.2&0.412&6.420&1.681&4.339&1.906&197.8, 177$^{\emph{c}}$\\
\end{tabular}
$^{\emph{a}}$ Reference \cite{LiuJAP}, $^{\emph{b}}$ Calculated
using present scheme with elastic constants at $T$=915$^{\circ}$C
from reference \cite{Heiming}, $^{\emph{c}}$ Reference
\cite{Heiming}.
\end{ruledtabular}
\end{table*}

In Table \ref{elastic}, we report our calculated lattice parameters,
bulk modulus, pressure derivative of the bulk modulus, and elastic
constants of $\alpha$, $\omega$, and $\beta$ Zr at different
pressures. For comparison, the experimental values
\cite{ZhaoPRB,Brandes,Barrett,LiuPRB,Heiming} and other theoretical
results \cite{HaoJPCM,Ikehata,Ahuja} are also listed. Obviously, our
calculated results accord well with experiments and other
calculations for $\alpha$ and $\omega$ phases. However, for $\beta$
Zr the difference of lattice parameter and bulk modulus between our
calculation and the experiment from Zhao \emph{et al.}
\cite{ZhaoPRB} is evident. This is due to the fact that the
experiment is conducted at \emph{T}=973 K, while our calculation is
valid only at 0 K. Similar temperature effect for $\beta$ Zr is also
found in studying the compression behaviour. As shown in Fig.
\ref{compression}, although good consistence with the experiment
\cite{Akahama2} at high pressure domain is clear, the values of the
atomic volumes are smaller than that by experiment at \emph{T}=973
K. Nevertheless, the compression performance of $\beta$ Zr is wholly
consistent with experiments. From Table \ref{elastic}, one can find
that the difference of elastic constants at zero pressure between
calculation and experiment is evident. This also can be attributed
to the temperature effect. In fact, the good accordance of the
lattice parameter and elastic constants of $\beta$ Zr at 0 GPa
between our calculation and previous DFT-PBE \cite{Ikehata} study
supplies the safeguard for our study of structure, mechanical, and
electronic properties of $\beta$ Zr. Additionally, the mechanical
stability of $\alpha$ and $\omega$ Zr at 0 GPa and at some typical
finite pressures can be predicted from the elastic constants data.
But the elastic constants of $\beta$ Zr at 0 GPa illustrate that the
cubic phase is mechanically unstable, which is in good agreement
with the results by Ahuja \emph{et al.} \cite{Ahuja} and Ikehata
\emph{et al.} \cite{Ikehata}. Actually, the mechanical stability of
$\beta$ Zr can be obtained through applying external pressure and
our data show that along with the increase of pressure from 0 GPa to
60 GPa, the value of $C_{11}-C_{12}$ increases almost linearly from
$-$5.7 GPa to 157 GPa. Fitting the curve (not shown) of the pressure
behaviour of $C_{11}-C_{12}$ by first-order polynomial function, we
find that the value of $C_{11}-C_{12}$ becomes positive at
$P\geq$3.13 GPa. In fact, at $P$=4 GPa our first-principles
calculated $C_{11}, C_{12}$, and $C_{44}$ equal to 101.8, 98.2, and
28.7 GPa, respectively, which explicitly indicate the elastic
stability of bcc Zr under this pressure.

After obtaining elastic constants, the elastic moduli, Poisson's
ratio ($\upsilon$), transverse sound velocities ($\upsilon_{t}$),
longitudinal sound velocities ($\upsilon_{l}$), average sound
velocities ($\upsilon_{m}$), and Debye temperature of $\alpha$,
$\omega$, and $\beta$ Zr at different pressures are calculated and
tabulated in Table \ref{mechanical}. The experimental values
\cite{LiuJAP,Heiming} are also collected in Table \ref{mechanical}
for comparison. The excellent coincidence between our calculation
and experiment for $\alpha$ and $\omega$ phases can be seen. For
$\beta$ phase, the bulk modulus \emph{B}, shear modulus \emph{G},
Young's modulus \emph{E}, and Debye temperature all linearly
increase with pressure. The Young's modulus \emph{E} at $P$=4 GPa
(just over the critical pressure 3.13 GPa for the elastic stability
of $\beta$ Zr) is calculated to equal to 31.97 GPa, which is very
near the Young's modulus value ($\mathtt{\sim}$30 GPa) of a human
bone. Thus, we successfully indicate the importance of stabilizing
the high temperature or pressure phase of Zr at ambient conditions.
In Table \ref{mechanical}, we also list the results of $\beta$ phase
calculated using present scheme with elastic and lattice constants
at $T$=915$^{\circ}$C from Ref. \cite{Heiming}. The derived Debye
temperature is slightly higher than that given in Ref.
\cite{Heiming}.

\begin{figure*}[ptb]
\begin{center}
\includegraphics[width=0.8\linewidth]{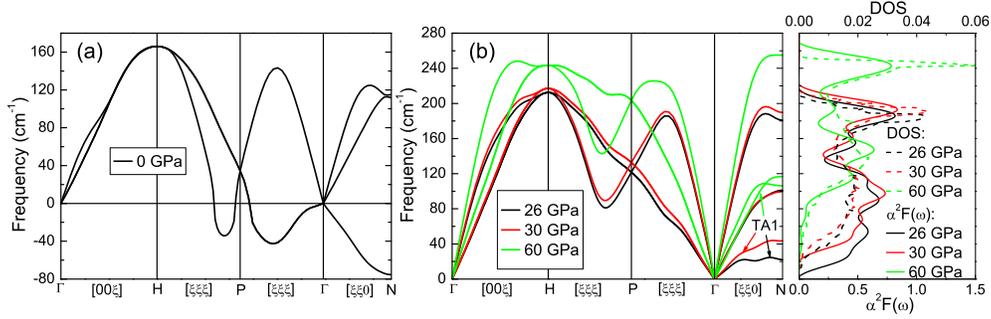}
\end{center}
\caption{(Color online) Phonon dispersion at 0 GPa (a) and phonon dispersion, Eliashberg phonon spectral function, and phonon DOS at 26, 30, and 60 GPa (b) for $\beta$ Zr.}%
\label{phonon}%
\end{figure*}

\begin{figure}[ptb]
\begin{center}
\includegraphics[width=0.5\linewidth]{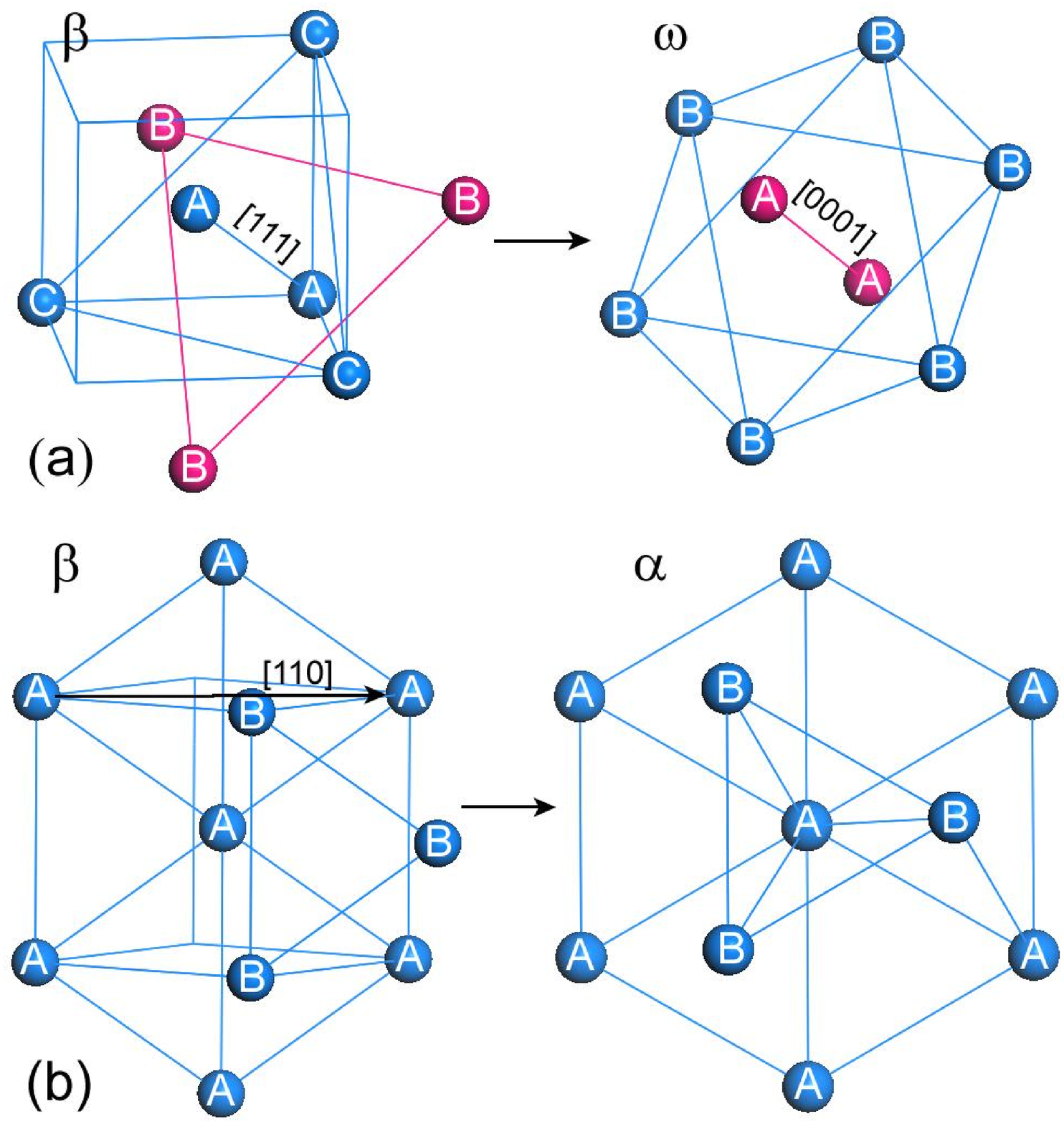}
\end{center}
\caption{(Color online) Schematic illustrations of the structural
transition for
(a) $\beta$$\mathtt{\rightarrow}$$\omega$ and (b) $\beta$$\mathtt{\rightarrow}$$\alpha$.}%
\label{path}%
\end{figure}

Phonon dispersion results for $\beta$ Zr at selected pressures are
shown in Fig. \ref{phonon}. Like the $\beta$ phase of Ti and Hf
\cite{Souvatzis,Mei}, the high pressure $\beta$ phase of Zr is also
dynamically unstable at low pressure [Fig. \ref{phonon}(a)]. Our
phonon dispersion calculations at high pressure illustrate that the
$\beta$ Zr becomes dynamically stable at around 26 GPa. In fact, no
imaginary frequencies are observed at 26-60 GPa pressure domain as
clearly indicated in Fig. \ref{phonon}(b). Therefore, although
recent experiment by P\'{e}rez-Prado and Zhilyaev \cite{Perez} has
successfully stabilized the high pressure/temperature phase of Zr at
ambient condition, our phonon dispersion calculations illustrate
that the $\beta$ phase is dynamically unstable in pressure domain of
0-25 GPa at 0 K. This fact confirms that the temperature effect on
the phase stability is also critical for the experimentally observed
\cite{Perez} $\beta$ phase at ambient condition. Besides, the
dynamical stability test at 26-60 GPa supplies the safeguard for our
following study of the superconductivity of $\beta$ Zr at high
pressure. As shown in Fig. \ref{phonon}(a), our phonon dispersion
results indicate soft modes for the longitudinal L
$\frac{2}{3}$[111] and transverse T $\frac{1}{2}$[110] phonons. As
indicated by experiment \cite{Heiming}, the L $\frac{2}{3}$[111]
unstable phonon branch is responsible for the
$\beta$$\mathtt{\rightarrow}$$\omega$ transition and the T
$\frac{1}{2}$[110] implies the $\beta$$\mathtt{\rightarrow}$$\alpha$
transition. In Fig. \ref{path}, we show the paths of
$\beta$$\mathtt{\rightarrow}$$\omega$ and
$\beta$$\mathtt{\rightarrow}$$\alpha$ transitions. As shown in Fig.
\ref{path}(a), the bcc $\beta$ phase can be viewed as ABC
periodically layered structure along the [111] direction and there
are two periods of ABC layers in its unit cell. For $\omega$ phase,
it can be viewed as AB periodical layered structure along the [0001]
direction [Fig. \ref{path}(a)]. The atoms on B and C layers in
$\beta$ phase move easily along the soft mode direction of [111] and
can transfer to be the B layer of $\omega$ phase. Therefore, the
atomic displacement along the [111] direction is the main origin for
the $\beta$$\mathtt{\rightarrow}$$\omega$ phase transition. For
$\beta$$\mathtt{\rightarrow}$$\alpha$ transition, one can regard the
$\beta$ phase as two layered (AB) structure along the [1$\bar{1}$0]
orientation [Fig. \ref{path}(b)]. In transition, firstly the atoms
on each layer expand along the [110] direction to form a hexagon,
and second, the adjacent (1$\bar{1}$0) planes slip relatively along
the [110] direction to create the hcp structure.

\begin{table}[ptb]
\caption{Calculated logarithmic average of vibrational frequencies
$\omega\rm{_{log}}$ and EPC constant $\lambda$ of $\alpha$,
$\omega$, and $\beta$ Zr at different pressures.}%
\label{superconductivity}
\begin{ruledtabular}
\begin{tabular}{cccccccccccccccccc}
Phase&Pressure&$\omega\rm{_{log}}$&$\lambda$\\
&(GPa)&(K)&\\
\hline
$\alpha$&0&142.52&0.8611\\
&5&139.26&0.8852\\
$\omega$&0&168.29&0.6723\\
&10&182.27&0.6753\\
&20&186.51&0.7302\\
&30&184.06&0.8174\\
$\beta$&26&80.91&2.5311\\
&28&123.43&2.0237\\
&30&136.22&1.7822\\
&40&169.17&1.2745\\
&50&198.83&1.0147\\
&60&221.55&0.8528\\
\end{tabular}
\end{ruledtabular}
\end{table}

\begin{figure}[ptb]
\begin{center}
\includegraphics[width=0.5\linewidth]{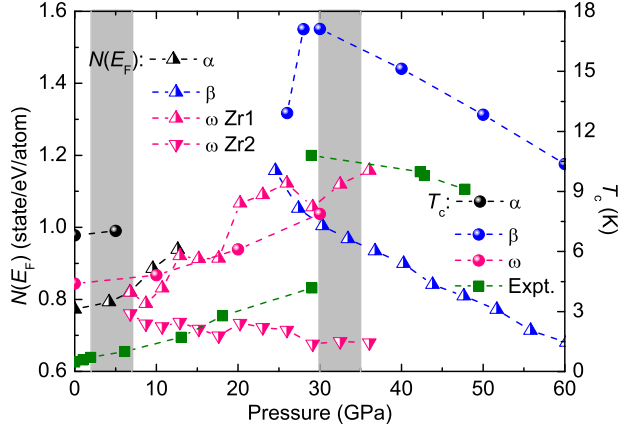}
\end{center}
\caption{(Color online) Calculated total density of state at the
Fermi level $N$(\emph{E}$_{\rm{F}}$) and superconducting transition
temperature $T$$_{c}$ as functions of pressure for $\alpha$,
$\omega$, and $\beta$ Zr. The experimental values of $T$$_{c}$ from
Ref. \cite{Akahama1} are also presented. The two shaded regions
indicate the $\alpha$$\mathtt{\rightarrow}$$\omega$ and
$\omega$$\mathtt{\rightarrow}$$\beta$ phase transition ranges of
2$-$7 GPa \cite{Sikka,Xia1,ZhaoPRB} and
30$-$35 GPa \cite{Xia1,Xia2,Akahama1,Akahama2,ZhaoAPL}, respectively.}%
\label{dos}%
\end{figure}

For high pressure phase of Zr, it has evident predominance for
high-temperature superconductivity. The EPC calculations have been
performed to explore the superconductivity. The Eliashberg phonon
spectral function $\alpha^{2}F(\omega)$ and phonon DOS at 26, 30,
and 60 GPa for $\beta$ Zr are shown in Fig. \ref{phonon}(b).
Logarithmic average phonon frequency $\omega\rm{_{log}}$ and the EPC
strength $\lambda$ of $\alpha$, $\omega$, and $\beta$ Zr at
different pressures are listed in Table \ref{superconductivity}. The
superconducting $\emph{T}_{\rm{c}}$ has been estimated from Eqs.
(3), and a typical value of Coulomb pseudopotential $\mu^{*}$=0.12
is used. Our calculated $\emph{T}_{\rm{c}}$, together with
$N$(\emph{E}$_{\rm{F}}$), as a function of pressure for the three
phases are plotted in Fig. \ref{dos}. Note that there are two
different kinds of atoms for $\omega$ Zr: Zr1 in 1$a$ and Zr2 in
2$d$ sites. Clearly, the predicted $\emph{T}_{\rm{c}}$ is in good
agreement with experimental results. As indicated by McMilan
equation [Eqs. (3)], $\emph{T}_{\rm{c}}$ has tight relationship with
$\omega\rm{_{log}}$, $\lambda$, and $N$(\emph{E}$_{\rm{F}}$). From
0-7 GPa, total $N$(\emph{E}$_{\rm{F}}$) of $\alpha$ phase increases
with pressure. In pressure range of 7-30 GPa, the average
$N$(\emph{E}$_{\rm{F}}$) of $\omega$ phase over two kinds of atomic
arrangements also exhibits increasing behavior upon compression. In
contrast, total $N$(\emph{E}$_{\rm{F}}$) of $\beta$ Zr decreases
with pressure in the whole domain of 26-60 GPa. All these
pressure-dependent behaviors of $N$(\emph{E}$_{\rm{F}}$) obey one
known fact that a more stable phase has a lower value of
$N$(\emph{E}$_{\rm{F}}$). Through plotting partial
$N$(\emph{E}$_{\rm{F}}$) of $s$, $p$, and $d$ orbitals (not shown),
we find that the main contribution to the pressure-dependent
behavior of the total $N$(\emph{E}$_{\rm{F}}$) comes from $d$
orbital. The $s$ orbital has no contribution and the contribution
from $p$ orbital is limited. From Fig. \ref{dos}, one can see that
the experimental $\emph{T}_{\rm{c}}$ changes in almost the same way
as that of the $N$(\emph{E}$_{\rm{F}}$). So, it seems that relative
larger value of the $N$(\emph{E}$_{\rm{F}}$) for $\beta$ Zr at 26
GPa may predict larger $\emph{T}_{\rm{c}}$ for superconductivity.
However, $N$(\emph{E}$_{\rm{F}}$) is less directly important than
$\omega\rm{_{log}}$ and $\lambda$ for $\emph{T}_{\rm{c}}$.

From Table \ref{superconductivity}, one can see that the calculated
$\lambda$ for $\alpha$ and $\omega$ phases increases with pressure
while $\omega\rm{_{log}}$ does not show any clear dependence on
pressure. As a result, the increasing behavior of
$\emph{T}_{\rm{c}}$ upon compression is mainly due to the increase
of $\lambda$. For $\beta$ Zr, $\omega\rm{_{log}}$ and $\lambda$ have
an almost opposite evolution with pressure. However, the changing
rate for $\omega\rm{_{log}}$ and $\lambda$ is different. The
increasing rate of $\omega\rm{_{log}}$ is clearly larger than the
decreasing rate of $\lambda$ in pressure range of 26 to 30 GPa.
After being compressed over 30 GPa, the decreasing rate of $\lambda$
exceeds the increasing rate of $\omega\rm{_{log}}$. So the main
affection to the pressure performance of $\emph{T}_{\rm{c}}$ comes
from the increase of $\omega\rm{_{log}}$ in pressure domain 26-30
GPa, while in 30-60 GPa is from the decrease of $\lambda$.
Therefore, the relatively small value of $\emph{T}_{\rm{c}}$ at 26
GPa for $\beta$ Zr is understandable. Furthermore, the lowest
transverse-acoustics (TA1) soft mode for bcc $\beta$ Zr in pressure
range of 26-30 GPa is evident [Fig. \ref{phonon}(b)] and can lead a
significant electron-phonon contribution. Therefore, as indicated in
Table \ref{superconductivity}, the EPC constant $\lambda$ at 26-30
GPa is predominantly larger than that of both $\alpha$ and $\omega$
phases. And, along with increasing the pressure, soft vibrational
mode for $\beta$ Zr gradually fade away, resulting in the decrease
of $\lambda$ and $\emph{T}_{\rm{c}}$. Thus, the decrease of
$\lambda$ and $\emph{T}_{\rm{c}}$ for $\beta$ Zr under pressure
mainly originates from the fading of soft vibrational modes induced
by increasing pressure.

\section{CONCLUSION}
In summary, the elasticity, dynamic properties, and
superconductivity of Zr under pressure up to 60 GPa have been
studied by means of the first-principles DFT-PBE method. Our results
have shown that the structural parameters, elastic constants,
elastic moduli, Poisson's ratio, ultrasonic velocities, and Debye
temperature of $\alpha$ and $\omega$ phases coincide well with
experiments. The $\alpha$$\mathtt{\rightarrow}$$\omega$ and
$\omega$$\mathtt{\rightarrow}$$\beta$ phase transition pressures at
$T$=0 K have been calculated to be $-$3.7 GPa and 32.4 GPa,
respectively. The compression curve of $\beta$ Zr is well consistent
with experiments. Our calculations have explicitly indicated that
the low elastic modulus of Zr occurs at low pressure in $\beta$
phase.

In phonon dispersion study, the soft modes of $\beta$ phase along
[111] and [110] directions have been shown and the
$\beta$$\mathtt{\rightarrow}$$\omega$ and
$\beta$$\mathtt{\rightarrow}$$\alpha$ transition paths have been
predicted. Dynamic stability test has illustrated that $\beta$ Zr is
unstable till being compressed over 25 GPa. In addition,
superconductivity has been obtained by electron-phonon coupling
calculations and our calculated $\emph{T}_{\rm{c}}$ accords well
with experiments. Through analyzing the electronic density of states
at the Fermi level, we have derived that the main contribution to
the pressure-dependent behavior of the superconductivity comes from
the $d$ orbital. The large $\emph{T}_{\rm{c}}$ at around 30 GPa for
$\beta$ Zr is mainly due to the TA1 soft mode. Under pressure, the
increase or decrease of $\emph{T}_{\rm{c}}$ for Zr in all three
phases has tight relation with the corresponding behavior of the EPC
constant $\lambda$.

\begin{acknowledgments}
This work was partically supported by NSFC under Grants No. 90921003
and No. 60776063, and by the Foundation for Development of Science
and Technology of China Academy of Engineering Physics under Grant
No. 2008B0101008 and No. 2009A0102005.
\end{acknowledgments}


\begin{thebibliography}{99}
\bibitem {Ikehata}H. Ikehata, N. Nagasako, T. Furuta, A. Fukumoto,
K. Miwa, and T. Saito, Phys. Rev. B \textbf{70}, 174113 (2004).

\bibitem {LiuPRB}W. Liu, B. Li, L. Wang, J. Zhang, and Y. Zhao,
Phys. Rev. B \textbf{76}, 144107 (2007).

\bibitem {Perez}M. T. P\'{e}rez-Prado and A. P. Zhilyaev, Phys.
Rev. Lett. \textbf{102}, 175504 (2009).

\bibitem {HuAPL}Q. M. Hu, S. J. Li, Y. L. Hao, R. Yang, B.
Johansson, and L. Vitos, Appl. Phys. Lett. \textbf{93}, 121902
(2008).

\bibitem {Sikka}S. K. Sikka, Y. K. Vohra, and R. Chidambaram, Prog.
Mater. Sci. \textbf{27}, 245 (1982).

\bibitem {Xia1}H. Xia, S. J. Duclos, A. L. Ruoff, and Y. K. Vohra,
Phys. Rev. Lett. \textbf{64}, 204 (1990).

\bibitem {ZhaoPRB}Y. S. Zhao, J. Z. Zhang, C. Pantea, J. Qian, L.
L. Daemen, P. A. Rigg, R. S. Hixson, G. T. Gray III, Y. P. Yang, L.
P. Wang, Y. B. Wang, and T. Y. Uchida, Phys. Rev. B \textbf{71},
184119 (2005).

\bibitem {Xia2}H. Xia, A. L. Ruoff, and Y. K. Vohra, Phys. Rev. B
\textbf{44}, 10374 (1991).

\bibitem {Akahama1}Y. Akahama, M. Kobayashi, and H. Kawamura, J.
Phys. Soc. Jpn. \textbf{59}, 3843 (1990).

\bibitem {Akahama2}Y. Akahama, M. Kobayashi, and H. Kawamura, J.
Phys. Soc. Jpn. \textbf{60}, 3211 (1991).

\bibitem {ZhaoAPL}Y. S. Zhao and J. Z. Zhang, Appl. Phys. Lett.
\textbf{91}, 201907 (2007).

\bibitem {PAW}P. E. Bl\"{o}chl, Phys. Rev. B \textbf{50}, 17953
(1994).

\bibitem {Kresse3}G. Kresse and J. Furthm\"{u}ller, Phys. Rev. B
\textbf{54}, 11169 (1996).

\bibitem {PBE}J. P. Perdew, K. Burke, and M. Ernzerhof, Phys. Rev.
Lett. \textbf{77}, 3865 (1996).

\bibitem {Monk}H. J. Monkhorst and J. D. Pack, Phys. Rev. B
\textbf{13}, 5188 (1972).

\bibitem {Baroni2}A. D. C. S. Baroni, S. de Gironcoli, P. Giannozzi, C. Cavazzoni,
G. Ballabio, S. Scandolo, G. Chiarotti, P. Focher, A. Pasquarello,
K. Laasonen, A. Trave, R. Car, N. Marzari, and A. Kokalj,
http://www.pwscf.org.

\bibitem {Baroni1}S. Baroni, P. Giannozzi, and A. Testa, Phys. Rev. Lett. \textbf{58}, 1861
(1987).

\bibitem {Giannozzi}P. Giannozzi, S. de Gironcoli, P. Pavone, and S. Baroni, Phys.
Rev. B \textbf{43}, 7231 (1991).

\bibitem {Birch}F. Birch, Phys. Rev. \textbf{71}, 809 (1947).

\bibitem {Hill}R. Hill, Phys. Phys. Soc. London \textbf{65}, 349
(1952).

\bibitem {Allen}P. B. Allen and R. C. Dynes, Phys. Rev. B \textbf{12}, 905 (1975).

\bibitem {McMillan}W. L. McMillan, Phys. Rev. \textbf{167}, 331 (1968).

\bibitem {HaoPRB}Y. J. Hao, L. Zhang, X. R. Chen, L. C. Cai, Q. Wu,
and D. Alf\`{e}, Phys. Rev. B \textbf{78}, 134101 (2008).

\bibitem {Schnell}I. Schnell and R. C. Albers, J. Phys.: Condens.
Matter \textbf{18}, 1483 (2006).

\bibitem {Barrett}C. S. Barrett and T. B. Massalski,
\emph{Structure of Metals} (New York: McGraw-Hill, 1966).

\bibitem {Heiming}A. Heiming, W. Petry, J. Trampenau, M. Alba, C.
Herzig, H. R. Schober, and G. Vogl, Phys. Rev. B \textbf{43}, 10948
(1991).

\bibitem {Brandes}E. A. Brandes, \emph{Smithells Metals Reference
Book} (Butterworth, London, 1983).

\bibitem {HaoJPCM}Y. J. Hao, L. Zhang, X. R. Chen, Y. H. Li, and H.
L. He, J. Phys.: Condens. Matter \textbf{20}, 235230 (2008).

\bibitem {Ahuja}R. Ahuja, J. M. Wills, B. Johansson, and O.
Eriksson, Phys. Rev. B \textbf{48}, 16269 (1993).

\bibitem {LiuJAP}W. Liu, B. Li, L. Wang, J. Zhang, and Y. Zhao, J.
Appl. Phys. \textbf{104}, 076102 (2008).

\bibitem {Souvatzis}P. Souvatzis, O. Eriksson, M. I. Katsnelson, and S. P. Rudin,
Phys. Rev. Lett. \textbf{100}, 095901 (2008).

\bibitem {Mei}Z. G. Mei, S. L. Shang, Y. Wang, and Z. K. Liu, Phys. Rev. B \textbf{80}, 104116 (2009).
\end{thebibliography}
\end{document}